\documentclass{PoS}%
\usepackage{graphicx}
\usepackage{amsmath}
\usepackage{amsfonts}
\usepackage{amssymb}%
\DeclareFontFamily{U}{rsfs}{\skewchar \font"7F}
\DeclareFontShape{U}{rsfs}{m}{n}{
	<-6> rsfs5
	<6-8> rsfs7
	<8-> rsfs10
	}{}
\DeclareMathAlphabet{\mathscr}{U}{rsfs}{m}{n}
\PoS{PoS(LAT2005)332}
\title{%
\begin{picture}(0,0)(0,0)%
\put(350,75){\makebox(0,0)[l]{%
\begin{minipage}{0.3\textwidth}
\normalsize
CHIBA-EP-156 \\ KEK Preprint 2005-66
\end{minipage}
}}%
\end{picture}%
Cho-Faddeev-Niemi decomposition of lattice Yang-Mills theory
and evidence of a novel magnetic condensation%
}

\ShortTitle{Cho-Faddeev-Niemi decomposition of lattice Yang-Mills theory
and ...}
\author{\speaker{Akihiro Shibata}\\
Computing Research Center, KEK, Tsukuba, 305-0801,Japan\\
E-mail:\email{akihiro.shibata@kek.jp}}
\author{Kei-Ichi Kondo\\
Department of Physics, Faculty of Science, Chiba University,
Chiba 263-8522, Japan \\
E-mail:\email{kondok@faculty.chiba-u.jp}}
\author{Seikou Kato \\
Takamatsu National College of Technology, Takamatsu 761-8058, Japan \\
E-mail:\email{kato@takamatsu-nct.ac.jp}}
\author{Takeharu Murakami and Toru Shinohara \\
Graduate School of Science and Technology,Chiba University,
Chiba 263-8522, Japan\\
E-mail:\email{tom@cuphd.nd.chiba-u.ac.jp},
\email{sinohara@cuphd.nd.chiba-u.ac.jp}}
\FullConference{XXIIIrd International Symposium on Lattice Field Theory\\
25-30 July 2005\\
Trinity College, Dublin, Ireland}
\abstract{
We present the first implementation of the Cho--Faddeev--Niemi decomposition
of the SU(2) Yang-Mills field on a lattice. Our construction retains the color
symmetry (global SU(2) gauge invariance) even after a new type of Maximally
Abelian gauge, as explicitly demonstrated by numerical simulations.%\cite{KKS05}
}
\begin{document}

\section{Introduction}

The Cho-Faddeev-Niemi (CFN) decomposition, or a change of variables of the
non-Abelian gauge potential in Yang--Mills theory, was proposed by Cho
\cite{Cho80} and Faddeev and Niemi \cite{FN98}. CFN decomposition introduces a
color vector field $\mathbf{n}(x)$ enabling us to extract explicitly the
magnetic monopole as a topological degree of freedom from the gauge potential
without introducing the fundamental scalar field in Yang--Mills theory. The
CFN decomposition has been formulated and extensively studied on the continuum
spacetime. For non-perturbative studies, however, it is desirable to put the
CFN decomposition on a lattice. This will enable us to perform powerful
numerical simulations to obtain fully non-perturbative results.

The main purpose of this paper is to propose a lattice formulation of the CFN
decomposition and to perform the numerical simulations on the lattice, paying
special attention to the magnetic condensations. Our lattice formulation
reflects a new viewpoint proposed by three of the authors in a previous paper
\cite{KMS05}, which enables us to retain the local and global gauge invariance
even after the new type of MAG. In the whole of this paper, we restrict the
gauge group to SU(2).

\section{CFN decomposition in the continuum}

We adopt the Cho-Faddeev-Niemi (CFN) decomposition for the non-Abelian gauge
field \cite{Cho80,FN98,Shabanov99}. By introducing a unit vector field
$\mathbf{n}(x)$ with three components, i.e., $\mathbf{n}(x)\cdot
\mathbf{n}(x):=n^{A}(x)n^{A}(x)=1$ $(A=1,2,3)$, the non-Abelian gauge field $%
%TCIMACRO{\TeXButton{A}{\mathscr{A}}}%
%BeginExpansion
\mathscr{A}%
%EndExpansion
_{\mu}(x)$ in the SU(2) Yang-Mills theory is decomposed as
\begin{equation}%
%TCIMACRO{\TeXButton{A}{\mathscr{A}}}%
%BeginExpansion
\mathscr{A}%
%EndExpansion
_{\mu}(x)=c_{\mu}(x)\mathbf{n}(x)+g^{-1}\partial_{\mu}\mathbf{n}%
(x)\times \mathbf{n}(x)+\mathbb{X}_{\mu}(x).
\end{equation}
We use the notation: $\mathbb{C}_{\mu}(x):=c_{\mu}(x)\mathbf{n}(x)$,
$\mathbb{B}_{\mu}(x):=g^{-1}\partial_{\mu}\mathbf{n}(x)\times \mathbf{n}(x)$
and $\mathbb{V}_{\mu}(x):=\mathbb{C}_{\mu}(x)+\mathbb{B}_{\mu}(x)$. By
definition, $\mathbb{C}_{\mu}(x)$ is parallel to $\mathbf{n}(x)$, while
$\mathbb{B}_{\mu}(x)$ is orthogonal to $\mathbf{n}(x)$. We require
$\mathbb{X}_{\mu}(x)$ to be orthogonal to $\mathbf{n}(x)$, i.e.,
$\mathbf{n}(x)\cdot \mathbb{X}_{\mu}(x)=0$. We call $\mathbb{C}_{\mu}(x)$ the
restricted potential, while $\mathbb{X}_{\mu}(x)$ is called the
gauge-covariant potential and $\mathbb{B}_{\mu}(x)$ is called the non-Abelian
magnetic potential. In the naive Abelian projection, $\mathbb{C}_{\mu}(x)$
corresponds to the diagonal component, while $\mathbb{X}_{\mu}(x)$ corresponds
to the off-diagonal component, apart from the vanishing magnetic part
$\mathbb{B}_{\mu}(x)$.

Accordingly, the non-Abelian field strength $%
%TCIMACRO{\TeXButton{F}{\mathscr{F}}}%
%BeginExpansion
\mathscr{F}%
%EndExpansion
_{\mu \nu}(x)$ is decomposed as
\begin{equation}%
%TCIMACRO{\TeXButton{F}{\mathscr{F}}}%
%BeginExpansion
\mathscr{F}%
%EndExpansion
_{\mu \nu}:=\partial_{\mu}%
%TCIMACRO{\TeXButton{A}{\mathscr{A}}}%
%BeginExpansion
\mathscr{A}%
%EndExpansion
_{\nu}-\partial_{\nu}%
%TCIMACRO{\TeXButton{A}{\mathscr{A}}}%
%BeginExpansion
\mathscr{A}%
%EndExpansion
_{\mu}+g%
%TCIMACRO{\TeXButton{A}{\mathscr{A}}}%
%BeginExpansion
\mathscr{A}%
%EndExpansion
_{\mu}\times%
%TCIMACRO{\TeXButton{A}{\mathscr{A}}}%
%BeginExpansion
\mathscr{A}%
%EndExpansion
_{\nu}=\mathbb{E}_{\mu \nu}+\mathbb{H}_{\mu \nu}+\hat{D}_{\mu}\mathbb{X}_{\nu
}-\hat{D}_{\nu}\mathbb{X}_{\mu}+g\mathbb{X}_{\mu}\times \mathbb{X}_{\nu},
\end{equation}
where we have introduced the covariant derivative in the background field
$\mathbb{V}_{\mu}$ by $\hat{D}_{\mu}[\mathbb{V}]\equiv \hat{D}_{\mu}%
:=\partial_{\mu}+g\mathbb{V}_{\mu}\times,$ and defined the two kinds of field
strength:
\begin{align}
\mathbb{E}_{\mu \nu}=  &  E_{\mu \nu}\mathbf{n},\quad E_{\mu \nu}:=\partial_{\mu
}c_{\nu}-\partial_{\nu}c_{\mu},\\
\mathbb{H}_{\mu \nu}=  &  H_{\mu \nu}\mathbf{n},H_{\mu \nu}:=-g^{-1}%
\mathbf{n}\cdot(\partial_{\mu}\mathbf{n}\times \partial_{\nu}\mathbf{n}).
\end{align}
We here notice that $\mathbb{G}_{\mu \nu}:=\mathbb{E}_{\mu \nu}+\mathbb{H}%
_{\mu \nu}$ is gauge invariant, while each of $\mathbb{E}_{\mu \nu}$ and
$\mathbb{H}_{\mu \nu}$ is not gauge invariant. We can define gauge invariant
monopole, called CFN monopole, using the field strength $\mathbb{G}_{\mu \nu}%
$\cite{KKS05}.

\section{CFN decomposition on a lattice}

We discuss how the CFN decomposition is implemented on a lattice by defining
the unit color vector field $\mathbf{n}_{x}$ to generate the ensemble of
$\mathbf{n}$-fields\cite{KKS05}. On lattice simulation, we can generate the
configurations of SU(2) link variables $\{U_{x,\mu}\}$, $U_{x,\mu}%
=\exp[-ig\epsilon%
%TCIMACRO{\TeXButton{A}{\mathscr{A}}}%
%BeginExpansion
\mathscr{A}%
%EndExpansion
_{\mu}(x)],$ using the standard Wilson action, where $\epsilon$ is the lattice
spacing and $g$ is the coupling constant. We use the continuum notation for
the Lie-algebra valued field variables, e.g., $%
%TCIMACRO{\TeXButton{A}{\mathscr{A}}}%
%BeginExpansion
\mathscr{A}%
%EndExpansion
_{\mu}(x)$, even on a lattice. To obtain the configuration with the same
Boltzmann weight as the original YM theory, we define CFN decomposition on a
lattice based on gauge symmetry of CFN variables.

\begin{figure}[tb]
\begin{center}
\begin{minipage}{0.47\textwidth}
\begin{center}
\fbox{\includegraphics[
height=1.2in,
width=2.1in
]%
{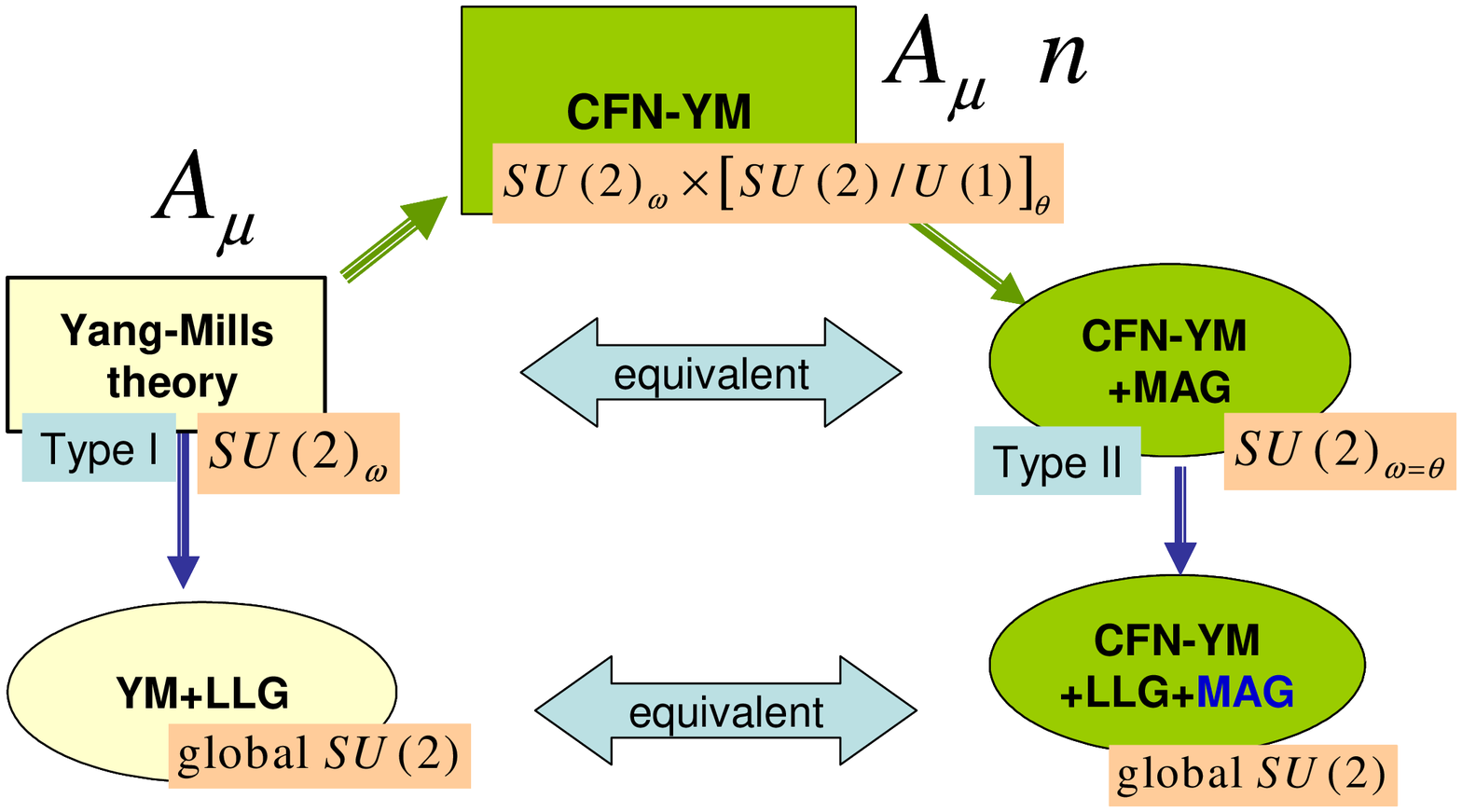}%
}\caption{Gauge symmetery of CFN-Yang-Mills theory }%
\label{fig:gaugesym}%
\end{center}
\end{minipage}
\hspace{5mm} \begin{minipage}{0.47\textwidth}
\begin{center}
\fbox{\includegraphics[
height=1.2in,
width=2.1in
]%
{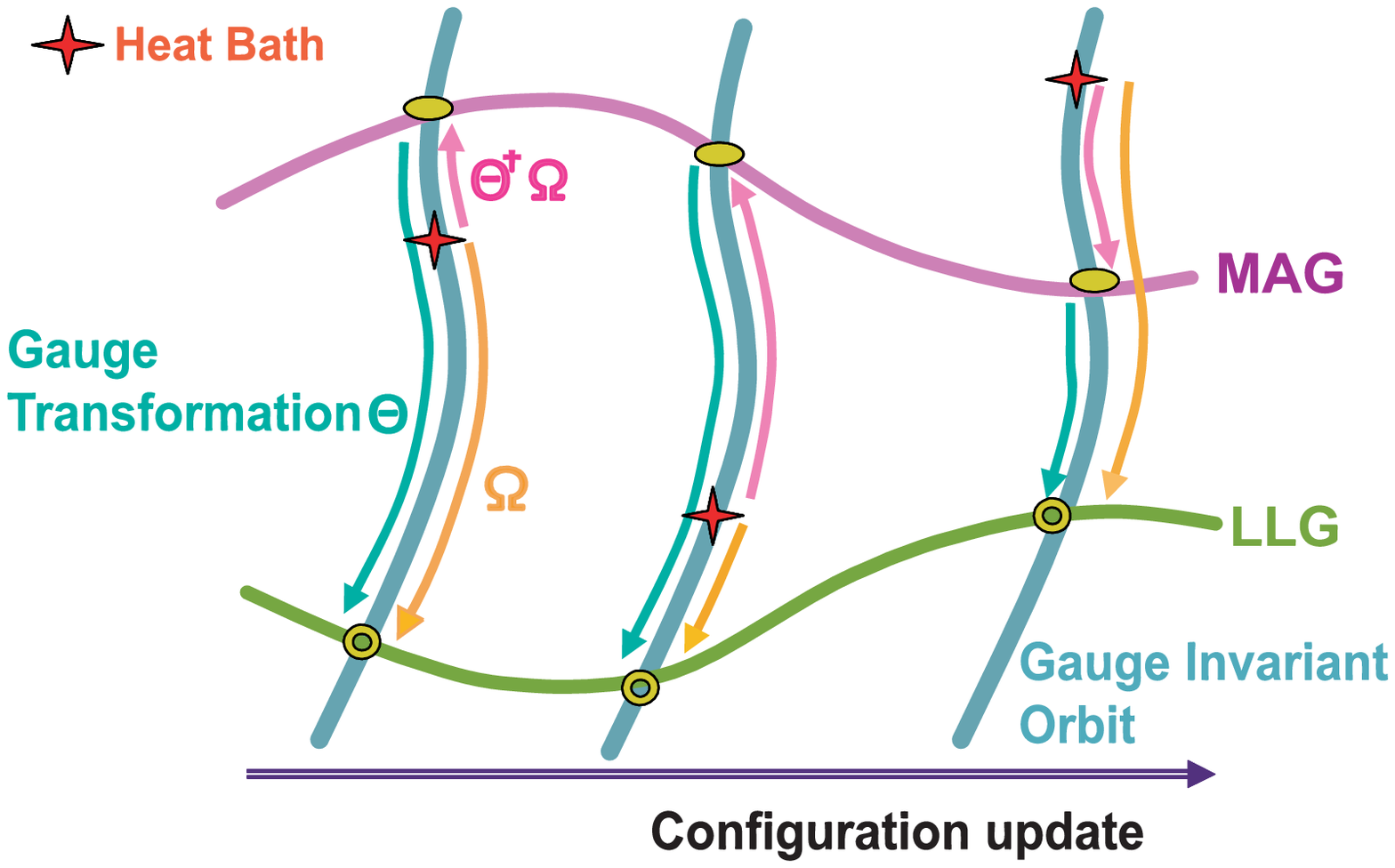}%
}\caption{Lattice CFN decomposition obtained by imposing nMAG and LLG}%
\label{fig:GF-orbit}%
\end{center}
\end{minipage}
\end{center}
\end{figure}

We define the CFN-Yang--Mills theory as the Yang-Mills theory written in terms
of the CFN variables. It has been shown \cite{KMS05} that the SU(2)
CFN-Yang--Mills theory has the local gauge symmetry $\tilde{G}_{local}%
^{\omega,\theta}:=SU(2)_{local}^{\omega}\times \lbrack SU(2)/U(1)]_{local}%
^{\theta}$ larger than one of the original Yang-Mills theory $SU(2)_{local}%
^{\omega}$, since we can rotate the CFN variable $\mathbf{n}(x)$ by angle
$\mathbf{\theta}^{\perp}(x)$ independently of the gauge transformation of $%
%TCIMACRO{\TeXButton{A}{\mathscr{A}}}%
%BeginExpansion
\mathscr{A}%
%EndExpansion
_{\mu}(x)$ by the parameter $\mathbf{\omega}(x)$. In order to fix the whole
enlarged local gauge symmetry $\tilde{G}_{local}^{\omega,\theta}$, we must
impose sufficient number of gauge fixing conditions. Recently, it has been
clarified \cite{KMS05} how the CFN-Yang--Mills theory can be equivalent to the
original Yang-Mills theory by imposing a new type of gauge fixing called the
new Maximal Abelian gauge (nMAG) to fix the extra local gauge invariance in
the continuum formulation, see Fig.\ref{fig:gaugesym}.

%%%%%%%%%%%%%%%%%%%%%%%%%%%%%%%%%%%%%%%%%%%%%%%%%%%%%%%%%%%%%%%%%%%%
%%%%%%%%%%%%%%%%%%%%%%%%%%%%%%%%%%%%%%%%%%%%%%%%%%%%%%%%%%%%%%%%%%%%%%%%%%

Corresponding to CFN Yang Mills theory in continuum, we define nMAG on a
lattice. By introducing a vector field $\mathbf{n}_{x}$ of a unit length with
three components, we consider a functional $F_{nMAG}[U,\mathbf{n}%
;\Omega,\Theta]$ written in terms of the gauge (link) variable $U_{x,\mu}$ and
the color (site) variable $\mathbf{n}_{x}$ defined by
\begin{equation}
F_{nMAG}[U,\mathbf{n};\Omega,\Theta]:=\sum_{x,\mu}\mathrm{tr}(\mathbf{1}%
-{}^{\Theta}\mathbf{n}_{x}{}^{\Omega}U_{x,\mu}{}^{\Theta}\mathbf{n}_{x+\mu}%
{}^{\Omega}U_{x,\mu}^{\dagger}).\label{eq:newMAGfun}%
\end{equation}
Here we have introduced the enlarged gauge transformation: ${}^{\Omega
}U_{x,\mu}:=\Omega_{x}U_{x,\mu}\Omega_{x+\mu}^{\dagger}$ for the link variable
$U_{x,\mu}$ and ${}^{\Theta}\mathbf{n}_{x}:=\Theta_{x}\mathbf{n}_{x}%
^{(0)}\Theta_{x}^{\dagger}$ for an initial site variable $\mathbf{n}_{x}%
^{(0)}$ where gauge group elements $\Omega_{x}$ and $\Theta_{x}$ are
independent SU(2) matrices on a site $x$. The former corresponds to the
$SU(2)^{\omega}$ gauge transformation $(%
%TCIMACRO{\TeXButton{A}{\mathscr{A}}}%
%BeginExpansion
\mathscr{A}%
%EndExpansion
_{\mu})^{\omega}(x)$ of the original potential, while the latter to the
adjoint $[SU(2)/U(1)]^{\theta}$ rotation, (see Fig.\ref{fig:gaugesym}).

After imposing the nMAG, the theory still has the local gauge symmetry
$SU(2)_{local}^{\omega=\theta}:=SU(2)_{local}^{II}$. Therefore, $\mathbf{n}%
_{x}$ configuration can not be determined at this stage. In order to
completely fix the gauge and determine $\mathbf{n}_{x}$, we need to impose
another gauge fixing condition for fixing $SU(2)_{local}^{II}$. In this paper
we choose the conventional Lorentz-Landau gauge or Lattice Landau gauge (LLG)
for this purpose. The LLG can be imposed by minimizing the function:
$F_{LLG}[U;\Omega]=\sum_{x,\mu}\mathrm{tr}(\mathbf{1}-{}^{\Omega}U_{x,\mu})$
with respect to the gauge transformation $\Omega_{x}$, (See
Fig.~\ref{fig:GF-orbit}). The LLG fixes the local gauge symmetry
$SU(2)_{local}^{\omega=\theta}=SU(2)_{local}^{II}$, while the LLG leaves the
global symmetry SU(2)$_{global}^{\omega}=SU(2)_{global}^{II}$ intact.
Therefore, we define the \textit{lattice nMAG} by minimizing the functional
$F_{nMAG}[U,\mathbf{n};\Omega,\Theta]$ with respect to the gauge
transformation $\{ \Omega_{x}\}$ and $\{ \Theta_{x}\}$, and we obtain the
\textit{lattice CFN decomposition}:%
\begin{equation}
{}^{\Omega}U_{x,\mu}=\exp \{-i\epsilon g[\mathbb{C}_{\mu}(x)+\mathbb{B}_{\mu
}(x)+\mathbb{X}_{\mu}(x)]\}.\label{LCFN}%
\end{equation}

Then a remaining issue to be clarified is how to construct the ensemble of the
color $\mathbf{n}$-fields used in defining $F_{nMAG}[U,\mathbf{n}%
;\Omega,\Theta]$ and $U_{x,\mu}$. Now it is shown that the desired color
vector field $\mathbf{n}_{x}$ is constructed from the interpolating gauge
transformation matrix $\Theta_{x}$ according to $\mathbf{n}_{x}^{(0)}%
=\sigma_{3}$ and
\begin{equation}
\mathbf{n}_{x}:=\Theta_{x}\sigma_{3}\Theta_{x}^{\dagger}=n_{x}^{A}\sigma
^{A},\quad n_{x}^{A}=\mathrm{tr}[\sigma_{A}\Theta_{x}\sigma_{3}\Theta
_{x}^{\dagger}]/2\quad(A=1,2,3).
\end{equation}
The first observation is that the functional $F_{nMAG}[U,\mathbf{n}%
;\Omega,\Theta]$ has another equivalent form $F_{nMAG}[U,\mathbf{n}%
;\Omega,\Theta]=F_{MAG}[U;G]=\sum_{x,\mu}\mathrm{tr}(\mathbf{1}-\sigma_{3}%
\ {}^{G}{}U_{x,\mu}\  \sigma_{3}\ {}^{G}{}U_{x,\mu}^{\dagger})$, with the
identification $G_{x}:=\Theta_{x}^{\dagger}\Omega_{x}$. This procedure
determines the configurations $\{G_{x}^{\ast}\}$ of SU(2) variables achieving
the minimum of $F_{MAG}[U;G],$ (See Fig.~\ref{fig:GF-orbit}).

On a gauge orbit, two representatives on the two gauge-fixing hypersurfaces
(MAG and LLG) are connected by the gauge transformation $\Theta \equiv
\Omega^{\ast}(G^{\ast})^{\dagger}$. Thus we can determine a set of
interpolating gauge-rotation matrices $\{ \Theta_{x}\}$ to construct
$\mathbf{n}_{x}$ ensemble. In fact, the color vector $\mathbf{n}_{x}$
constructed in this way represents a real-valued vector $\vec{n}_{x}%
=(n_{x}^{1},n_{x}^{2},n_{x}^{3})$ of unit length with three components, and
transforms in the adjoint representation under the gauge transformation II.
%, i.e.,
%$\bm{n}_{x} \rightarrow \Omega{}'{}_{x} \bm{n}_{x} \Omega{}'{}_{x}^\dagger$
%or $\delta_{\omega}' \bm{n}_{x}= g \bm{n}_{x} \times \bm{\omega}{}'{}_{x}$
%with $\Omega{}'{}_{x}=e^{ig \bm{\omega}{}'_x}$ ($\bm{\omega}{}'_x:= \bm{\omega}_x=\bm{\theta}_x$).

By imposing simultaneously the nMAG and the LLG in this way, we can completely
fix the whole local gauge invariance $\tilde{G}_{local}^{\omega,\theta}$ of
the lattice CFN-Yang--Mills theory. It should be remarked that, even after the
gauge fixing, the global (color) symmetry $SU(2)_{global}^{\omega=\theta}$ is unbroken.

Finally, we give the explicit expressions of CFN\ variables on lattice, from
link variable $U$ and $\mathbf{n}$-fields. In the continuum theory, the
CFN\ decomposition is uniquely obtained as $c_{\mu}(x)=\mathbf{n}%
(x)\mathbf{\cdot}%
%TCIMACRO{\TeXButton{A}{\mathscr{A}}}%
%BeginExpansion
\mathscr{A}%
%EndExpansion
_{\mu}(x)$ and\ $\mathbb{X}_{\mu}=g^{-1}\mathbf{n}(x)\mathbf{\times}\hat
{D}_{\mu}[%
%TCIMACRO{\TeXButton{A}{\mathscr{A}}}%
%BeginExpansion
\mathscr{A}%
%EndExpansion
_{\mu}]\mathbf{n}(x)$ for the given gauge potential $\
%TCIMACRO{\TeXButton{A}{\mathscr{A}}}%
%BeginExpansion
\mathscr{A}%
%EndExpansion
_{\mu}(x)$ and $\mathbf{n}$-field \cite{KKMSS05}. \ Thus CFN\ variables on a
lattice can be defined by the gauge potential $g\epsilon \mathbb{A}_{x,\mu
}:=\frac{i}{2}\left(  U_{x,\mu}-U_{x,\mu}^{\dagger}\right)  ,$ using U-linear
definition. We define the differential of $n$-field as a forwarded
differential corresponding to a link variable, $\Delta_{\mu}\vec{n}%
(x):=\vec{n}(x+\epsilon \hat{\mu})-\gamma_{\mu}(x)\vec{n}(x)$, where $\hat{\mu
}$ is a unit vector to the direction $\mu$. The coefficient $\gamma_{\mu
}(x)=\vec{n}(x+\epsilon \hat{\mu})\cdot \vec{n}(x)$ is defined to satisfy the
orthogonal condition $\Delta_{\mu}\vec{n}(x)\cdot \vec{n}(x)=0$ ($\partial
_{\mu}\vec{n}(x)\cdot \vec{n}(x)=0$ in continuum theory). So the CFN variables
on a lattice are obtained as follows;
\begin{align}
\qquad \mathbb{B}_{x,\mu} &  :=\Delta_{\mu}\vec{n}_{x}\times \vec{n}_{x}=\vec
{n}_{x+\epsilon \hat{\mu}}\times \vec{n}_{x},\\
c_{x,\mu} &  =\mathbb{A}_{x,\mu}\cdot \mathbf{n}_{x},\qquad \mathbb{X}_{x,\mu
}:=\mathbb{A}_{x,\mu}-c_{x,\mu}\mathbf{n}_{x}-\mathbb{B}_{x,\mu}.
\end{align}

\section{Numerical simulations}

Our numerical simulations are performed as follows. In the continuum
formulation, the CFN variables were introduced as a change of variables in
such a way that they do not break the global gauge symmetry $SU(2)_{global}%
^{II}$ or "color symmetry", corresponding to the global gauge symmetry
$SU(2)_{global}^{I}$ in the original Yang-Mills theory. Hence the nMAG can be
imposed in terms of the CFN variables without breaking the color symmetry.
This is a crucial difference between the nMAG based on the CFN decomposition
and the conventional MAG based on the ordinary Cartan decomposition which
breaks the $SU(2)_{global}$ explicitly. Therefore, we must perform the
numerical simulations so as to preserve the color symmetry as much as
possible.\footnote{Whether the color symmetry is spontaneously broken or not
is another issue to be investigated separately.} This is in fact possible as follows.

Remember that the nMAG on a lattice is achieved by repeatedly performing the
gauge transformations. In order to preserve the global SU(2) symmetry, we
adopt a random (global) gauge transformation only in the first sweep among the
whole sweeps of gauge transformations in the standard iterative gauge fixing
procedure for the LLG. This procedure moves an ensemble of unit vectors
$\mathbf{n}_{x}$ to a random ensemble of $\mathbf{n}_{x}$ which is far away
from $\mathbf{n}_{x}=(0,0,1)$. Then we search for the local minima around this
configuration of $\mathbf{n}_{x}$ by performing the successive gauge
transformations. The first random gauge transformation as well as the
subsequent gauge transformations are accumulated to obtain the gauge
transformation matrix $\Theta$ by which $\mathbf{n}$ is constructed.

\subsection{Lattice data}

Our numerical simulations are performed on the lattice with the lattice size
$L^{4}=$ $24^{4}$ by using heat-bath method for the standard Wilson
action\cite{Creutz80} for the gauge coupling $\beta=2.3\sim2.7$ and periodic
boundary conditions. Staring with cold initial condition and thermalizing
50*100 sweeps, we have obtained 200 configurations at intervals of 100 sweeps.
For LLG and MAG, we have used the over relaxation algorithm.

\begin{figure}[tb]
\begin{center}
\begin{minipage}{0.48\textwidth}
\begin{center}
\includegraphics[ height=1.6in, width=2.36in]
{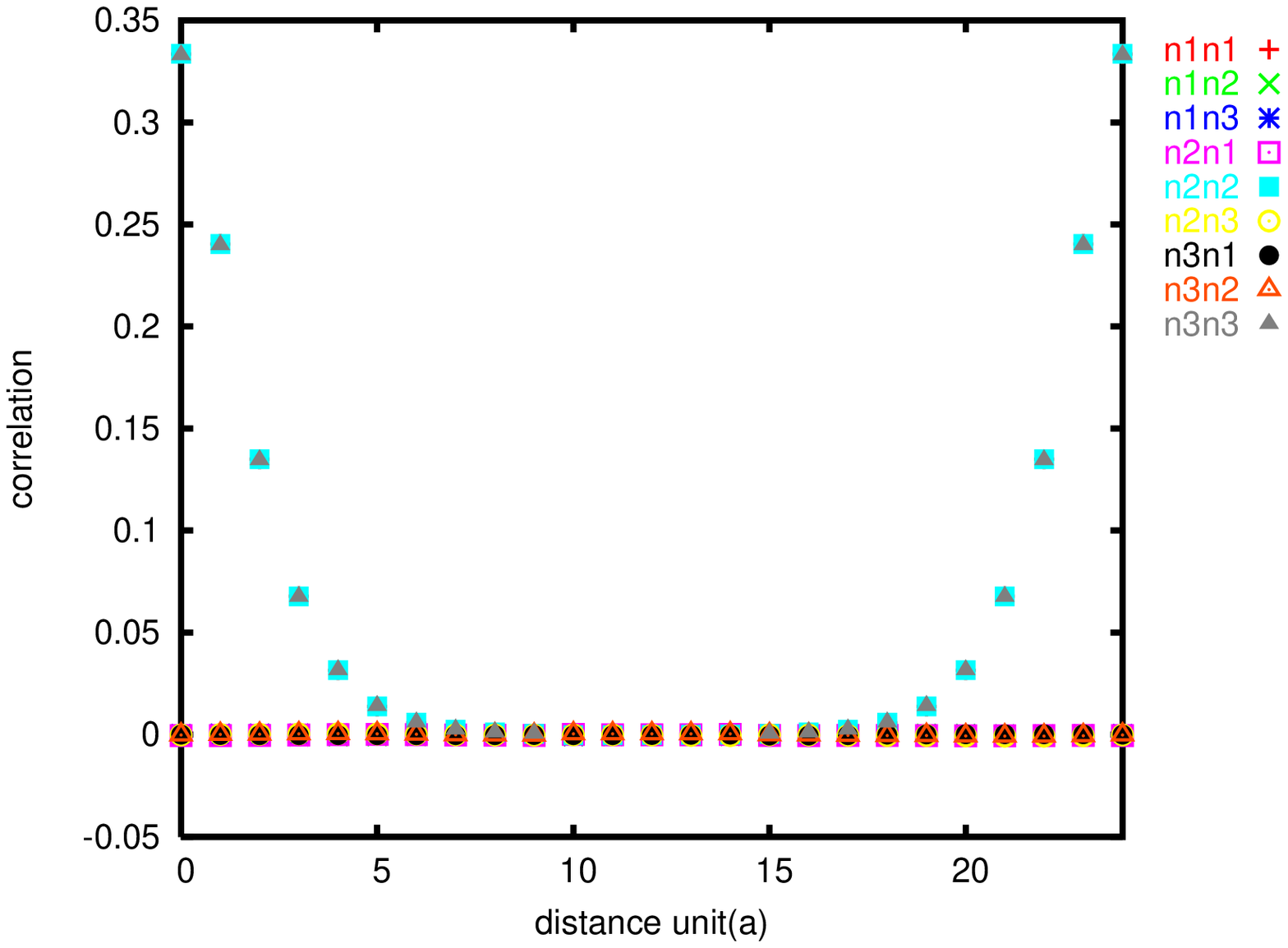}
\caption{<nn> correlation}%
\label{fig:nn-coll}%
\end{center}
\end{minipage} \begin{minipage}{0.48\textwidth}
\begin{center}
\includegraphics[height=1.6in,width=2.36in]%
{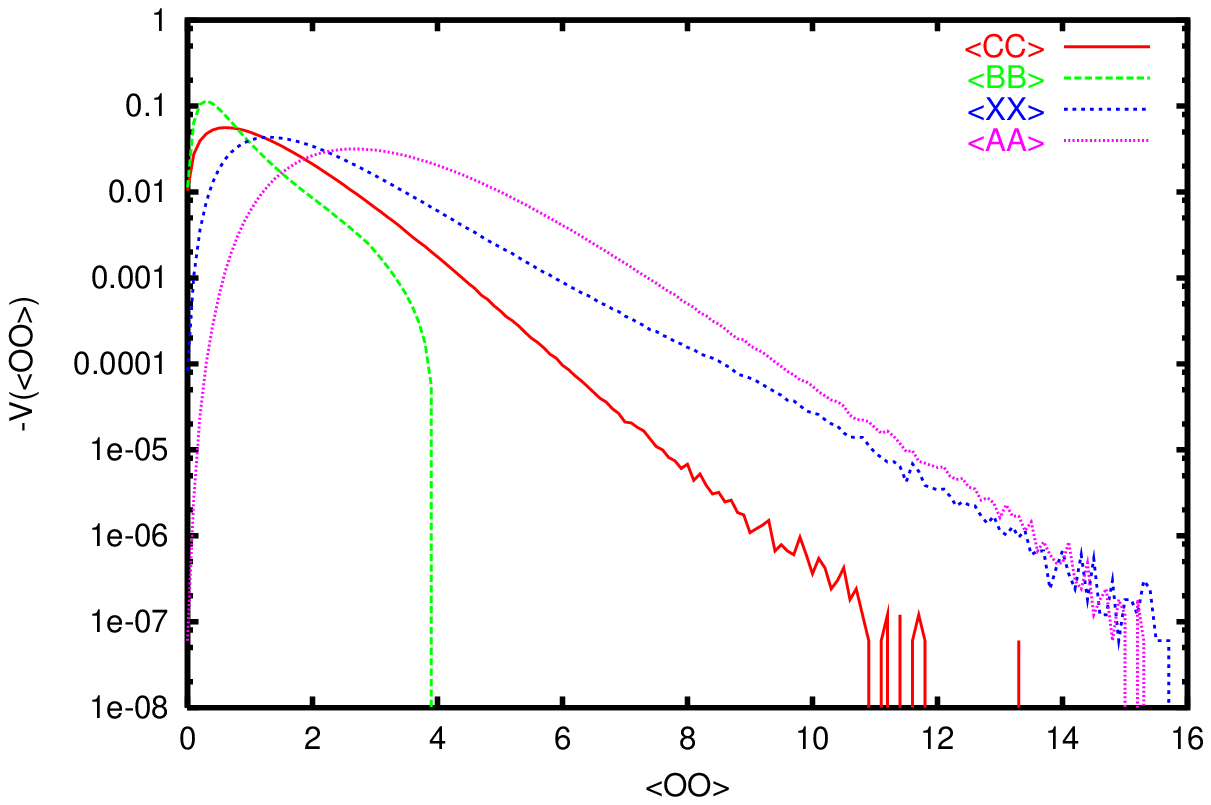}%
\caption{constraint effective potential}
\label{fig:effv}
\end{center}
\end{minipage}
\end{center}
\end{figure}

We first focus on the $\mathbf{n}$-fields. We have measured expectation values
$\left \langle n_{x}^{A}\right \rangle $ $(A=1,2,3)$, and have obtained the
vanishing expectation values. Moreover, we have measured the two-point
correlation functions defined by $\left \langle n_{x}^{A}n_{0}^{B}\right \rangle
$, see Fig.~\ref{fig:nn-coll}. The two-point correlation functions
$\left \langle n_{x}^{A}n_{0}^{A}\right \rangle $ (no summation over $A$)
exhibit almost the same behavior in all the directions ($A=1,2,3$), while
$\left \langle n_{x}^{A}n_{0}^{B}\right \rangle $ ($A\not =B$) vanish. Thus, we
have obtained the correlation function $\left \langle n_{x}^{A}n_{0}%
^{B}\right \rangle =\delta^{AB}D(x)$ respecting color symmetry. These results
indicate that \textit{the global $SU(2)$ symmetry (color symmetry) is unbroken
in our main simulations}.

Secondary, we focus on the anatomy of the condensation by CFN variables. We
measure the "naive" condensation of mass dimension 2; $\left(  g\epsilon
\right)  ^{2}\left \langle \mathbb{A}_{x,\mu}^{2}\right \rangle =\left(
g\epsilon \right)  ^{2}\left[  \left \langle c_{x,\mu}^{2}\right \rangle
+\left \langle \mathbb{B}_{x,\mu}^{2}\right \rangle +\left \langle \mathbb{X}%
_{x,\mu}^{2}\right \rangle +2\left \langle \mathbb{B}_{x,\mu}\mathbb{X}_{x,\mu
}\right \rangle \right]  $, and have obtained the non-zero value of
condensaions of CFN\ variables.  We also measure the "constraint effective
potential" defined by probability distribution of a local operator $\Phi(x);$
Pr$(\phi)=\left \langle \delta \left(  \phi(x)-\Phi(x)\right)  \right \rangle $
and the effective potential is defined by $V_{eff}(\phi)=-\ln \left \langle
\delta \left(  \phi(x)-\Phi(x)\right)  \right \rangle .$ Here, we measure
effective potential for $\Phi(x)=\left \{  \mathbb{A}_{x,\mu}^{2},\text{
}c_{x,\mu}^{2},\text{ }\mathbb{B}_{x,\mu}^{2},\text{ }\mathbb{X}_{x,\mu}%
^{2},...\right \}  $. Fig. \ref{fig:effv} shows that effective potential
$-V_{eff}(\phi)$ has peaks at nonzero expectation value $\left \langle
\Phi(x)\right \rangle .$ This suggest that we can expect non-zero condensations.

\section{Summary and Discussion}

We have shown how to implement the CFN decomposition (change of variables) of
SU(2) Yang--Mills theory on a lattice, according to a new viewpoint proposed
in \cite{KMS05}. A remarkable point is that our approach can preserve both the
local SU(2) gauge symmetry and the color symmetry (global SU(2) symmetry) even
after imposing a new type of gauge fixing (called the nMAG) which is regarded
as a constraint to reproduce the original Yang-Mills theory.

Moreover, we have succeeded to perform the numerical simulations in such a way
that the color symmetry is unbroken. This is the first remarkable result. This
is in sharp contrast to the previous approaches \cite{DHW02,Shabanov01}.
Although the similar technique of constructing the unit vector field
$\mathbf{n}_{x}$ from a $SU(2)$ matrix $\Theta$ has already appeared, e.g., in
\cite{DHW02,Shabanov01,IS00}, there is a crucial conceptual difference between
our approach and others.

We have shown preliminary result for the anatomy of the condensation of mass
dimension 2. We have obtained the result suggesting non-zero value of the
condensations. To determine the physical value of the condensation, we need
the further study such as the renomalization of the CFN variables.

\section*{Acknowledgment}

The numerical simulations have been done on a supercomputer (NEC SX-5) at
Research Center for Nuclear Physics (RCNP), Osaka University. This project is
also supported in part by the Large Scale Simulation Program No.133 (FY2005)
of High Energy Accelerator Research Organization (KEK). This work is
financially supported by Grant-in-Aid for Scientific Research (C)14540243 from
Japan Society for the Promotion of Science (JSPS), and in part by Grant-in-Aid
for Scientific Research on Priority Areas (B)13135203 from the Ministry of
Education, Culture, Sports, Science and Technology (MEXT).


\begin{thebibliography}{99}                                                                                               %


\bibitem {Cho80}Y.M. Cho,
%Restricted gauge theory,
Phys. Rev. D \textbf{21}, 1080
%--
(1980).
%\\
%Y.M. Cho,
%Extended gauge theory and its mass spectrum,
Phys. Rev. D \textbf{23}, 2415
%--
(1981).

\bibitem {FN98}L. Faddeev and A.J. Niemi,
%Partially dual variables in SU(2) Yang-Mills theory,
[hep-th/9807069], Phys. Rev. Lett. \textbf{82}, 1624
%--
(1999).

\bibitem {KKS05}S. Kato, K.-I. Kondo, T. Murakami, A. Shibata and T.
Shinohara, and S. Ito, hep-lat/0509069

\bibitem {KMS05}K.-I. Kondo, T. Murakami and T. Shinohara,
%Yang--Mills theory constracted from Cho--Faddeev--Niemi decomposition,
%Preprint CHIBA-EP-151,
hep-th/0504107.

\bibitem {Kondo04}K.-I. Kondo,
%MAGNETIC CONDENSATION, ABELIAN DOMINANCE, AND INSTABILITY OF SAVVIDY VACUUM,
[hep-th/0404252], Phys.Lett. B \textbf{600}, 287
%--296
(2004).

\bibitem {KKMSS05}S. Kato, K.-I. Kondo, T. Murakami, A. Shibata and T.
Shinohara,
%Numerical evidence for the existence of a novel magnetic condensation in Yang-Mills theory,
%Preprint, CHIBA-EP-150/KEK Preprint 2005-6,
hep-ph/0504054.

\bibitem {Shabanov99}S.V. Shabanov,
%An effective action for monopoles and knot solitons in Yang-Mills theory,
[hep-th/9903223], Phys. Lett. B \textbf{458}, 322
%--330
(1999). \newline S.V. Shabanov,
%Yang-Mills theory as an Abelian theory without gauge fixing,
[hep-th/9907182], Phys. Lett. B \textbf{463}, 263
%--272
(1999).

\bibitem {Creutz80}M. Creutz,
%Monte Carlo study of quantized SU(2) gauge theory,
Phys. Rev. D\textbf{21}, 2308
%--2315
(1980).

\bibitem {Shabanov01}S.V. Shabanov,
%Infrared Yang-Mills theory as a spin system,
[hep-lat/0110065], Phys. Lett. B \textbf{522}, 201
%--209
(2001).

\bibitem {DHW02}L. Dittmann, T. Heinzl and A. Wipf,
%Effective sigma models and lattice Ward identities,
[hep-lat/0210021], JHEP\textbf{0212}, 014 (2002).

\bibitem {IS00}H. Ichie and H. Suganuma,
%Monopoles and gluon fields in QCD in the maximally abelian gauge,
[hep-lat/9808054], Nucl. Phys. B \textbf{574}, 70
%--106
(2000).
\end{thebibliography}
\end{document}